\documentclass[prl,twocolumn,showpacs,superscriptaddress]{revtex4}
\usepackage[dvips]{graphicx}
\usepackage{dcolumn}
\usepackage{bm}
\usepackage{amssymb}
\usepackage{textcomp}

\DeclareGraphicsExtensions{eps}

\begin{document}
\preprint{APS/123-QED}
\title{Unconventional Dynamics in Triangular Heisenberg Antiferromagnet NaCrO$_2$}
\author{A. Olariu}
\author{P. Mendels}
\author{F. Bert}
\affiliation{%
Laboratoire de Physique des Solides, UMR 8502 CNRS, Universit\'{e}
Paris-Sud, 91405 Orsay, France}%
\author{B. G. Ueland}
\affiliation{%
Department of Physics, Pennsylvania State University, University Park, Pennsylvania 16802, USA  }%
\author{P. Schiffer}
\affiliation{%
Department of Physics, Pennsylvania State University, University Park, Pennsylvania 16802, USA  }%
\author{R. F. Berger}
\affiliation{%
Department of Chemistry, Princeton University, Princeton, New Jersey 08540, USA }%
\author{R. J. Cava}
\affiliation{%
Department of Chemistry, Princeton University, Princeton, New Jersey 08540, USA }%

\date{October 18, 2006}

\begin{abstract}
We report magnetization, specific heat, muon spin rotation, and Na
NMR measurements on the $S=3/2$ rhombohedrally stacked Heisenberg
antiferromagnet NaCrO$_2$. This compound appears to be a good
candidate for the study of isotropic triangular Heisenberg
antiferromagnets with very weak interlayer coupling. While specific
heat and magnetization measurements indicate the onset of a
transition in the range $T_c\sim~40-50~K$, both muon spin rotation
and NMR reveal a fluctuating crossover regime extending well below
$T_c$, with a peak of relaxation rate $T_1^{-1}$ around
$T\approx25~K$. This novel finding is discussed within the context
of excitations in the triangular Heisenberg antiferromagnets.
\end{abstract}

\maketitle

Since the initial proposal by Anderson for a liquid-like resonating
valence bond state in triangular $S=1/2$ antiferromagnets
(AF)~\cite{Anderson73}, a huge number of novel states have been
unveiled in triangle-based lattices. The frustration generated by
the geometry of the lattice is responsible for specific low-energy
excitations which contribute to destabilization of any possible
magnetic ordering. In the past decade, the fluctuating character at
very low temperatures has become a prominent feature of experimental
realizations of corner sharing, e.g. kagome or pyrochlore lattices.
By contrast, earlier studies on edge-sharing stacked triangular AF
displayed spin ordering with remarkable $H-T$ phase diagrams, and
the issue of new universality classes was
addressed~\cite{Collins97}. There are strong indications that the
concept of chirality, i.e., the way the spins are rotated in a
$120^{\circ}$ N\'eel order for a given triangle, might be important
with, possibly, an original set of low-energy
excitations~\cite{Kawamura98}. Notably, the systems which have
received the most attention consist of anisotropic
spins~\cite{Collins97}, and the canonical Heisenberg case with
isotropic interactions has not been a subject of many experimental
studies.

The field of triangular Heisenberg antiferromagnets was revived
quite recently both by the unpredicted discovery of
superconductivity in the stacked triangular Na cobaltates family
Na$_x$CoO$_2$ and the spin liquidlike behavior observed in
triangular compounds down to very low $T$
~\cite{Nakatsuji05,Shimizu03,Coldea01}. In Na$_x$CoO$_2$, a very
rich phase diagram was discovered, which may combine many possibly
competing parameters such as charge order, magnetic frustration, and
strong electronic correlations. In this context, the isostructural
insulating magnetic compound NaCrO$_2$ ($S=3/2$) appears to be an
ideal candidate to study the effects of frustration in
isolation~\cite{footnote1}. The well separated Cr$^{3+}$ planes
stack in a strict rhombohedral \emph{R$\bar{3}$m ABCABC} structure,
implying a single value of the in-plane Cr-Cr exchange constant. In
the very limited data available on this compound, the maximum of the
magnetic susceptibility $\chi$ was taken as an indication of a
transition around 49~K~\cite{Delmas78}, and early neutron scattering
work~\cite{Soubeyroux79} demonstrated a 2D character to the magnetic
correlations. EPR measurements~\cite{Elliston75,Angelov84} yield a
relative $g$ anisotropy of 0.25\% and point to a small single-ion
uniaxial $c$ anisotropy $\lesssim$~1~K, therefore a dominant
Heisenberg character. Finally, data on the more studied LiCrO$_2$
member of this family~\cite{Delmas78,Kadowaki95,Angelov84} have
revealed 120$^{\circ}$ spin-ordering in a plane containing the
$c$-axis, which was tentatively attributed to a small easy-axis
anisotropy.

In this paper, we present the first detailed local study of
NaCrO$_2$. Through NMR and muon spin rotation ($\mu$SR), we have
both investigated the static magnetic properties and revealed the
existence of an intermediate dynamical fluctuating regime well below
40~K, a behavior which has not been previously observed in studies
of triangular Heisenberg antiferromagnets (THAF).

Polycrystalline NaCrO$_2$ was prepared by mixing high purity
Na$_2$CO$_3$ and Cr$_2$O$_3$, with 2\% Na in excess of the
stoichiometric amount, and grinding in an agate mortar. The mixture
was pressed into pellets, wrapped in Zr foil, and heated in a dense
alumina boat at $T=750^{\circ}C$ for 30 h under flowing Ar, with one
intermediate grinding. Characterization of the samples by powder
x-ray diffraction showed them to be single phase with excellent
crystalline quality.

dc magnetic susceptibility measurements performed in the $T=5-300~K$
range (Fig.~1) are in perfect agreement with an earlier study up to
$T=800~K$, and yield $\Theta_{CW}\sim~290~K$ and $\mu_{eff}$=3.78
$\mu_{B}$~\cite{Delmas78}. Previous work suggested that the in-plane
antiferromagnetic exchange $J\sim~20~K$ was due mainly to direct
exchange through the overlap of nn Cr$^{3+}$
orbitals~\cite{Motida70}. We have found that $\chi$ continuously
increases down to $T=49~K$, where a broad maximum is observed. From
a comparison with samples having Ga substituted for Cr, the low-$T$
increase of $\chi$ can be attributed to less than 1\% spinless
defects.

We also measured specific heat using a Quantum Design PPMS system.
Measurements on the isostructural, nonmagnetic compound NaScO$_2$
were scaled and used to subtract out lattice contributions. The
resulting magnetic specific heat is displayed in Fig.~1. The peak at
$T_c=41~K$ is much broader than would be expected for a long range
ordering transition.

\begin{figure}
\includegraphics[width=8cm]{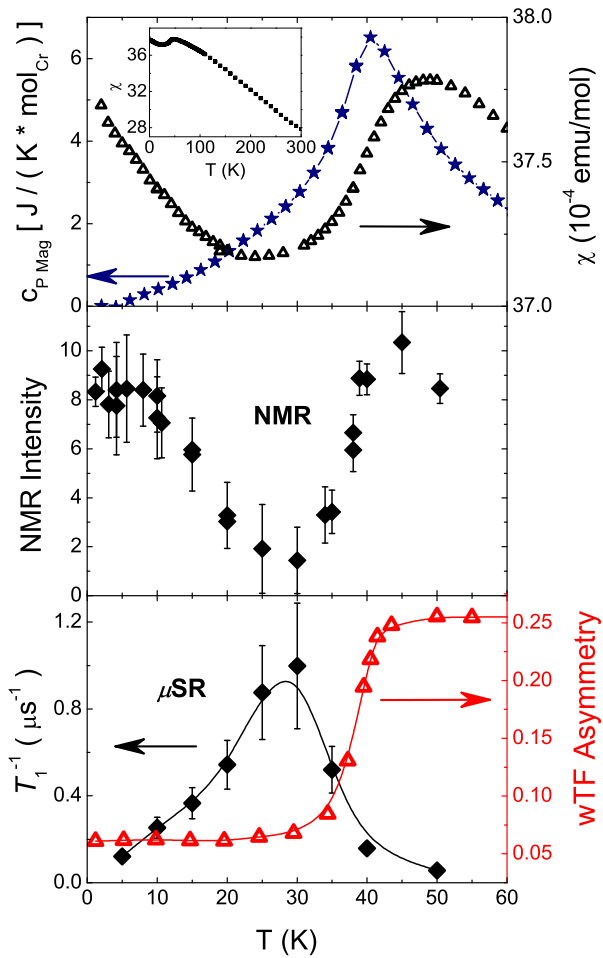}
\caption{Top: Magnetic specific heat (left axis) and magnetic
susceptibility (right axis) versus temperature. Inset: $\chi(T)$ on
an expanded $T$ range. Middle panel: Evolution of the NMR intensity
with $T$. Bottom: $\mu$SR, left: $T$ variation of $T_1^{-1}$, right:
weak transverse field $\mu$SR asymmetry (see text).}
\end{figure}

\begin{figure}
\includegraphics[width=9cm]{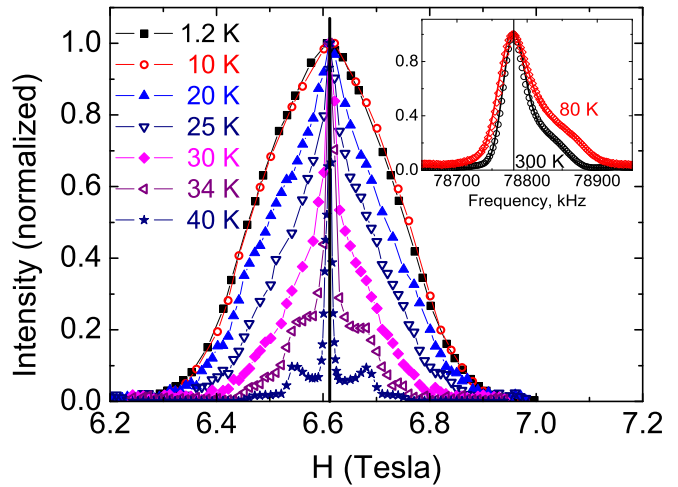}
\caption{$^{23}$Na NMR spectra for $T<80~K$. The vertical bar
represents the unshifted reference field. The spectra have been
normalized to the maximum intensity. First order quadrupolar
singularities are clearly observed at 40~K for $H=6.54~T$ and
6.68~T. Inset: Fourier transform spectra at 80~K and 300~K with
dipolar fits (continuous lines, see text).}
\end{figure}

Na-NMR measurements were carried out in the range $T=1.2-300$~K. The
$^{23}$Na nucleus ($\gamma /2\pi=11.2618$~MHz/T) has a spin
$I$~=~3/2 and is located in between the center of two Cr triangles
from adjacent layers. Above $T=80$~K, the spectra were obtained by
combining spin echo Fourier transforms taken at various frequencies
in a $H=7$~T fixed field. Below $T=80$~K, the spectra were recorded
by sweeping the magnetic field in the $H=6-7$~T range at a fixed
frequency of 74.527~MHz.

We first focus on the $-1/2\rightarrow1/2$ transition in the
paramagnetic regime (Fig.~2, inset). The line shift hardly varies
(max 0.007\%) between $T=80-300$~K, whereas a sizeable linewidth of
approximately 0.1\%, slightly varying with temperature, is observed.
The hyperfine contribution to the coupling between Cr and Na
(mediated by the oxygen) is therefore very weak and dominated by the
Cr$^{3+}$-Na dipolar interaction. This is quantitatively confirmed
by our dipolar simulations, which account for the observed
asymmetric lineshape (see Fig.~2, inset), and agree within 20\% with
the Cr$^{3+}$ magnetic moment extracted from our SQUID measurements.
As a comparison, the hyperfine coupling is here 70 times smaller
than in Na$_{0.7}$CoO$_2$. Since Na-O paths are only slightly
different between these compounds, we can attribute the smallness of
the Na-O-Cr hyperfine coupling to a negligible overlap of O with Cr,
which is expected from the more ionic character of Cr. Hence, we can
safely infer that (i) the intralayer Cr-O-Cr coupling is dominated
by \emph{direct} Cr-Cr exchange, which confirms
Ref.~\cite{Motida70}, and (ii) the exchange between Cr in different
layers is negligible.  Therefore the interlayer Cr-Cr coupling is
found to be of a dipolar nature, and thus very weak ($\sim$~0.01~K).
This indicates that NaCrO$_2$ is an excellent experimental
realization of a THAF.

We now describe the low-$T$ NMR local study of the frozen state
(Fig.~2). In the case of a conventional magnetic transition at
$T_c=41$~K, one would expect a growth of the internal field,
\emph{i.e.} the order parameter, within a few K below $T_c$. Because
of powder averaging, this should translate into a drastic broadening
of the NMR spectrum. Strikingly, we find that the broadening is
evident only below 30~K, much lower than $T_c$. As an illustration,
one can still distinguish the first order quadrupolar singularities
above 30~K, and only a negligible broadening of the central line
occurs for 30$<$T$<$40~K. At lower $T$, the linewidth at half
maximum of the central line is found to saturate below 10~K
(Fig.~3).

Because of powder averaging, each Na site inequivalent with respect
to the magnetic structure will contribute to the lineshape through a
rectangular pattern, the width of which is twice the internal field
$H_{int}$. In a simple conventional $120^{\circ}$ three-lattice
N\'{e}el state, a given Cr plane yields two sets of Na sites,
associated with the alternating $\pm$ chirality of adjacent Cr
triangles. The number of overall inequivalent sites, and, thus, the
distribution of $H_{int}$ depend drastically on the type of stacking
along the $c$ axis. In addition, dipolar field values strongly
depend on the orientations of the moments. Therefore, it was found
to be unfeasible to simply discriminate from the low-$T$ dome-shaped
line between various plausible ordered
structures~\cite{Soubeyroux79,Kadowaki95}. It is also not possible
to distinguish between a disordered freezing and a quite complex
ordered state. In any case, the low-$T$ saturation value of $1500$~G
determined from the linewidth is  consistent with a simple estimate
of the dipolar field using one nearby plane only. A precise
knowledge of the magnetic structure would be necessary for any
further quantitative discussion.

The major finding from our NMR study lies in the wipe-out of the NMR
integrated intensity (Fig.~1), \emph{i.e.} the number of detected
nuclei~\cite{footnote2}. The loss of intensity results from a
distribution of relaxation times, a large part of which becomes
shorter than the $\sim$~10~$\mu$s lower limit of the NMR time
window~\cite{Hunt99,Mendels00,MacLaughlin76}. In conventional
transitions, the wipe out of the signal occurs over a very narrow
$T$ range around the transition and is related to the slowing down
of fluctuations from the paramagnetic to the static state. Here the
full intensity is only recovered below 10~K, indicating a 30~K-broad
regime of slow fluctuations. Also the minimum intensity, \emph{i.e.}
the maximum of relaxation, is around $T=25(5)$~K, at least 25\%
lower than $T_c$ where the first signs of freezing are observed.
This original \emph{intermediate extended fluctuating regime} is
most likely related to the frustrated triangular network of
NaCrO$_2$.

\begin{figure}
\includegraphics[width=9cm]{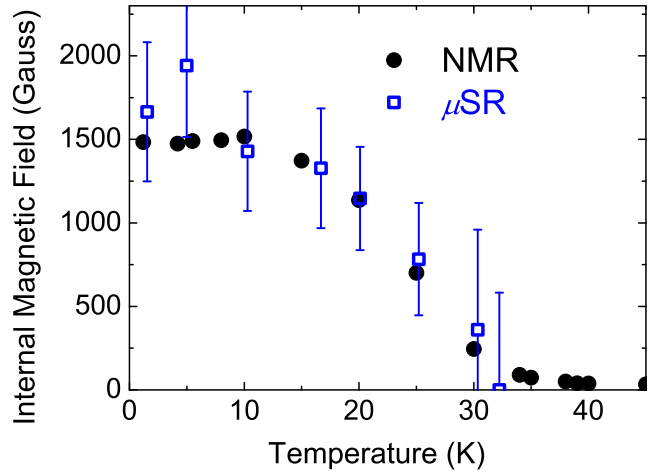}
\caption{$T$-evolution of the internal field measured at the site of
the \emph{detected} nuclei or of the muon (see text).}
\end{figure}

In order to get more insight into this dynamical regime, we
performed complementary $\mu$SR experiments at PSI (Switzerland) and
ISIS (UK) facilities. The $\mu$SR time window (10~ns~-~15~$\mu$s) is
much better suited to tracking the persisting dynamics in slowly
fluctuating magnets. In addition, $\mu$SR allows us to probe
\emph{all} sites, whereas only a weak fraction of the sites were
detected in NMR between 20 and 40~K. Fully polarized implanted
$\mu^+$ (gyromagnetic ratio $\gamma_{\mu}/2\pi=135.5$~MHz/T)
interact through dipolar coupling with the local magnetic
environment. As commonly assumed in oxides, muons most probably stop
$\sim$~1~\AA $ $ away from O$^{2-}$ sites.

Weak transverse field (20~G) experiments were performed to track the
spin freezing, since only muons close to paramagnetic (unfrozen)
sites should oscillate around the applied field direction. Figure~1
shows a drastic loss of asymmetry below 40~K, which indicates a
freezing corresponding to the specific heat maximum.

\begin{figure}
\includegraphics[width=9cm]{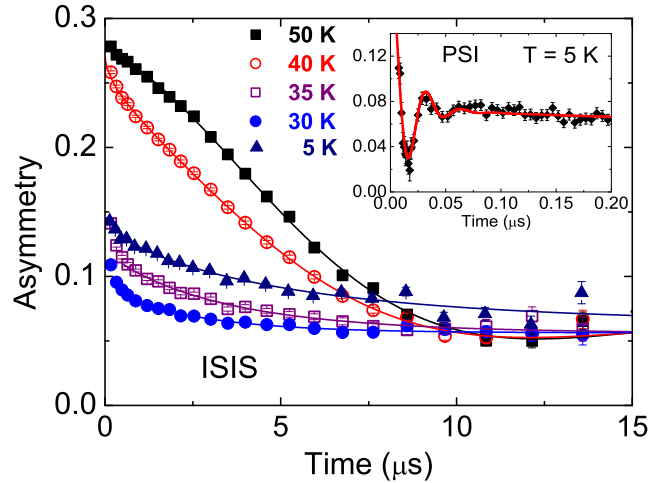}
\caption{Main panel (ISIS): $\mu$SR asymmetry in zero external
field. Inset (PSI): Early times asymmetry at $T=5~K$. The background
is different for the two setups. Lines are for fits.}
\end{figure}

Zero field asymmetry curves are presented in Fig.~4. Above $T_c$, in
the fast fluctuation regime, muons mainly sense a weak magnetic
field due to $\sim$~2~G static nuclear dipoles. The decrease of the
$\mu^+$ polarization had the expected Kubo-Toyabe shape, Gaussian at
early times, whereas electronic spin fluctuations result in very
modest relaxation with
$T_1^{-1}\sim$~0.06~$\mu$s$^{-1}$~\cite{footnote3}.

Below the transition temperature, we expect the asymmetry to evolve
with time according to:
\begin{equation}
A=A_0 [ \frac{2}{3} cos (2\pi\nu t+\phi)e^{-\lambda
t}+\frac{1}{3}]e^{-(t/T_1)^{\alpha}}+B
\end{equation}
The constant $A_0$, of the order of 0.25, stands for the maximum
asymmetry at 100\% beam polarization, and $B$ arises from muons not
stopping inside the sample. The oscillating term corresponds to the
$\mu^+$ precession around the internal magnetic field ($H_{\mu}$)
direction at the average frequency $\nu=\gamma_{\mu}H_{\mu}$, while
the damping of the oscillation is due to the width of the internal
field distribution $\lambda/\gamma_{\mu}$. In the case of static
behavior, we obtained the so called "one-third tail" at long times,
$A\to A_0/3$+B, since we expect $T_1^{-1}$ to be negligibly small.
In general, dynamical processes can be clearly singled out at long
times on the relaxation of this one-third tail, which enables one to
keep track of the evolution of $T_{1}^{-1}$ with $T$.

A high statistics run taken at $T=5$~K is displayed in the inset in
Fig.~4. The early time oscillating behavior is made clear by the two
visible wiggles and rules out a completely disordered picture. A
$H_\mu\sim$~1900~G internal field is estimated from the
$\nu=$~26~MHz frequency of the oscillations. The fast early time
damping indicates a large 1600~G distribution of $H_\mu$, maybe
related to sizeable disorder. The weakness of relaxation indicates a
purely static magnetic frozen phase in the low-$T$ limit.

At higher $T$, the internal field is found to decrease and cannot be
followed any more above 32~K (Fig.~3). The order of magnitude of the
internal field agrees very well with our NMR findings, issued in
both cases from dipolar coupling of the probe to its surrounding.

We now focus on the relaxation effects evidenced on the $t>0.1~\mu$s
one third tail displayed in the main frame in Fig.~4. The increase
of relaxation between 5 and 30~K is clearly visible,
$T_1^{-1}$(30~K)$\sim$1~$\mu$s$^{-1}$, followed by a decrease at
$T=35$~K (Fig.~1, bottom). A broad maximum of the relaxation rate is
found to occur around 25-30~K~\cite{footnote4}. This peak was found
to be progressively washed out in samples having Ga substituted for
Cr, which demonstrated its intrinsic character.

To summarize, our specific heat, susceptibility, NMR wipe-out and
$\mu$SR data point to an \emph{onset} of gradual freezing at
$T_c\approx$~41~K, associated with a slowing down of spin
fluctuations. Below this temperature, a crossover regime occurs,
marked by a maximum in the relaxation rate at 0.75~$T_c$. This broad
fluctuating regime extends down to 0.25~$T_c$. Neither a
conventional transition at 41~K nor around 30~K can simply reconcile
these findings.

One possibility is that the $T_1^{-1}$ peak signals a subsequent
transition. To our knowledge, the splitting of the transition
temperature has been observed only in a very narrow temperature
range ($<$3~K) for triangular AF, e.g. the quasi-2D VX$_2$ (X~=~Cl,
Br, I) and the quasi-1D ABX$_3$ systems~\cite{Collins97}, both with
single-ion anisotropy. The scenario of a second magnetic transition
at $T=30$~K can be safely ruled out here in view of the weak
anisotropy, which contrasts with the large temperature range between
$T_c$ and the $T_1^{-1}$ maximum. Also, no sign of a second peak is
detected in the specific heat, and finally, the broad fluctuating
regime points to a dynamical crossover between states rather than a
sharp transition. Whatever the transition scenario, our study
definitely reveals novel excitations. Such a specific feature of the
isotropic THAF was clearly pointed out by recent
calculations~\cite{Zheng06,Fujimoto06}.

One can speculate that the dynamical regime could be evidence for
topological "Z$_2$" excitations in the chirality vortex
regime~\cite{Kawamura98}. The latter can be viewed as the pendant of
excitations for the XY spins, well known to drive the
Kosterlitz-Thouless transition. The 30~K temperature could then be
the equivalent of that transition, i.e., resulting from the coupling
between the Z$_2$ vortex and anti-vortex. Interestingly, the
existence of Z$_2$ vortices has been suggested above $T_c$ to
interpret EPR data in LiCrO$_2$ and VX$_2$~\cite{Ajiro88,Kojima93}
and for the spin liquid state found in
NiGa$_2$S$_4$~\cite{Nakatsuji05}. Whether the spin texture of the
ground state can be singled out is a matter for future research.

Regardless of interpretation, with its 2D Heisenberg character,
NaCrO$_2$ altogether with recent
systems~\cite{Nakatsuji05,Shimizu03,Coldea01} form a bridge between
two intensively studied classes of systems. On one side, corner
sharing Heisenberg AF, characterized by a fluctuating regime
extending down to very low $T$ and marginal orders, e.g. the kagome
based compounds~\cite{Mendels00,Bono04,Bert05}. On the other side,
the case of anisotropic triangular AF, which order at low $T$ with
two successive transitions in a narrow $T$ range. Our data strongly
support renewed study of the ACrO$_2$ family including local
measurements of the A~=~Li, K compounds and modern neutron studies
of the entire family.

We wish to thank H. Alloul, J. Bobroff, O. C\'{e}pas, M. Elhajal, G.
Lang, L. L\'{e}vy, C. Lhuillier, P. Viot for fruitful discussions
and A. D. Hillier and A. Amato for assistance at $\mu$SR facilities.
This work was supported by the European Commission EC FP6 grants. We
gratefully acknowledge support from NSF grant DMR-0353610, CNRS-NSF
and I2CAM (B. U.).



\begin{thebibliography}{19}
\bibitem{footnote1}
\bibinfo{author}{In the parent NaCoO$_2$ compound Co$^{3+}$ is
nonmagnetic.}
\bibitem{footnote2}
\bibinfo{author}{In our experiment we use a conventional
$\pi/2-\tau-\pi$ pulse sequence. The detected signal was
extrapolated to $\tau\to$0 to calculate the intensity and corrected
for $T$ effects.}
\bibitem{footnote3}
\bibinfo{author}{Early times on high-$T$ spectra (not shown) reveal about 5\% spurious magnetic phase.}
\bibitem{footnote4}
\bibinfo{author}{Note that the value of the
stretched exponent $\alpha\sim 0.5$ indicates a distribution either
of $T_1$ or of muon sites.}
\end{thebibliography}

\end{document}